
\magnification=\magstep1
\hsize=5.5 true in

\centerline{ ON THE OCCURRENCE OF NAKED SINGULARITY IN}
\centerline{ SPHERICALLY SYMMETRIC GRAVITATIONAL COLLAPSE$^{+}$}
\vfill
\centerline{\bf I. H. Dwivedi$^{*}$ and P. S. Joshi  }
\centerline{\bf Tata Institute of Fundamental Research}
\centerline{\bf Homi Bhabha Road, Bombay 400 005}
\centerline{\bf India}
\vfill
\noindent {\bf $^{*}$ Permanent Address}:\hfill\break
\noindent{\bf Institute of Basic Sciences}\hfill\break
\noindent{\bf Agra University}\hfill\break
\noindent{\bf Khandari, Agra, India}\hfill\break
\bigskip

$^{+}$ To appear in Commun. Math. Physics (1994)

\vfil\eject

\magnification=\magstep1  
\hoffset=0 true cm        
\hsize=6.0 true in        
\vsize=8.5 true in        
\baselineskip=24 true pt plus 0.1 pt minus 0.1 pt 
\overfullrule=0pt         
\input tap

\centerline{Abstract}

Generalizing earlier results of Joshi and Dwivedi (Commun. Math. Phys.
146, 333 (1992); Lett. Math. Phys. 27, 235 (1993)), we analyze here
the spherically symmetric gravitational collapse of a matter cloud
with a general form of matter for the formation of a naked
singularity. It is shown that this is related basically
to the choice of initial data to the Einstein field equations,
and would therefore occur in generic situations from  regular initial
data within
the general context considered here, subject to the matter satisfying
the weak energy condition. The condition on initial data which leads
to the formation of black hole is also characterized.

\vfil\eject

\beginsection{ 1. INTRODUCTION}

We considered recently [1] the formation and structure of naked
singularity in the self-similar gravitational collapse of a perfect
fluid with an
adiabatic equation of state, and also for a general form of matter
subject only to the weak energy condition but with an arbitrary equation
of state. It was shown in those cases that strong curvature naked
singularities form in the gravitational collapse from a regular initial
data, from which non-zero
measure families of non-spacelike trajectories could come out.
The criterion
for the existence of such singularities was characterized in terms of
the existence of real positive roots of an algebraic equation constructed
out of the field variables.

The considerations such as those in [1] and [2] provide many
insights into the phenomena of gravitational collapse. For example,
the Einstein equations, under the geometric assumption of self-similarity
reduce to ordinary differential equations.
This allows one to construct explicit collapse scenarios such as
the Vaidya-Papapetrou radiation collapse with a linear mass function [3],
or perfect fluid collapse [1,4] which provide useful information on the
phenomena of gravitational collapse. It is also known that such
conclusions are not restricted to self-similar spacetimes only [2,5].

The scenarios such as above have, however, limitations.
For example, while in [1] the equation of state is general, subject to
the weak energy condition only, the geometric assumption of
self-similarity
is there. On the other hand, while [2] considers a more general
collapse, the classes considered there have other limitations.
For example, the later reference there is restricted to dust collapse
(however, with a generic mass function with only $C^1$ differentiability)
whereas the former, while considering a wide class of matter
(with the condition that the mass function be expandable about the
singularity)
excludes some important collapse situations due to
some other assumptions
made there (e.g. the metric coefficient $c(v,r)$ is also expandable
about the singularity and $c(v,0)=1$).
The presently known collapse scenarios are restricted mainly to dust
and perfect fluid.
While the form of matter such as a perfect fluid has
wide range of physical applications with the advantage of incorporating
the pressure which could be important in the later stages of collapse,
it is certainly important to examine if similar conclusions will hold
for  other reasonable forms of matter.
For example, as pointed out by
Eardley [6], dust could be an approximation to a
more fundamental form of matter, such as a massive scalar field.
It is thus possible that the naked singularity is an
artifact of the approximation used, and not a basic feature of
collapse. It is  therefore important to consider the collapse
phenomena
for a rather general form of matter, without limitations such as
above. This should help us to understand
gravitational collapse and the
occurrence of naked singularity in a more clear manner, which should
lead to a more precise formulation of the cosmic censorship
hypothesis.

Our purpose here is to analyze the formation of naked singularities
in spherically symmetric collapse from this perspective for a general
form of matter, only subject to the weak energy condition
with no restriction on the equation of state.
All the presently known naked singular examples,
such as radiation collapse,
dust or perfect fluid models could apply only to a rather narrow class
of equations of state. Our considerations here show, apart from other
implications, that given any equation of state, for which there does
exist a spherically symmetric, naked singularity example of present type,
then for all sufficiently close equations of state (suitably defined)
there is also such an example. We reduce the spherically symmetric
Einstein field equations to a single parabolic partial differential
equation of second order. The class of naked singular spacetimes with
such equations of state is defined in terms of solutions of this
equation.

In the following, we therefore analyze the formation of naked
singularities within a broader framework as mentioned above.
In section 2 the basic equations for the collapse
are set up and section 3 discusses the initial value problem to consider
the gravitational collapse of a spherically symmetric matter
cloud which is initially non-singular. The existence of naked singularity
is characterized in section 4, also examining its curvature strength.
It is pointed out that the occurrence of naked singularity or a black hole
is more a problem of the choice of the initial data for the field equations
rather than the form of matter or the equation of state.
The concluding section 5 briefly considers the implications.

\beginsection {2. SPHERICALLY SYMMETRIC COLLAPSE}

We consider here the final fate of collapse of
a matter cloud that evolves from
a regular physical data defined on an initial spacelike surface.
The energy-momentum tensor has a compact support on this initial
surface where all the physical quantities such as density etc. are
regular and finite. For sufficiently high mass, there is no
stability configuration possible and the collapse results into a
space-time singularity as implied by the singularity theorems in
general relativity. This singularity is characterized by the
existence of a future directed non-spacelike trajectory in the matter
cloud which is future incomplete, having a finite
affine length but no future end point.

To consider a general matter field, we note that
the  stress-energy tensor $T^a_b$ describing the
matter distribution of a space-time
can be classified as being one of the following four types [7].
These are the possibilities when it has either
in a timelike invariant 2-plane
(i) two real orthogonal eignvectors,
(ii) one double null real eignvector,
(iii) no real eignvector,
or  has in null invariant 2-plane,
(iv) one triple null real eignvector.
The matter distributions of type (iii) and (iv)
necessarily violate the energy conditions ensuring the positivity of the
mass-energy density.
Furthermore, such fields have not been observed so far.
Thus, we would not attribute any physical interpretation to the same
presently. The only observed occurrence of
type (ii) matter distribution  corresponds
to  zero rest mass fields representing
directed radiation. In a spherically symmetric
space-time these
could be effectively described
by the Vaidya metric. The radiation collapse,
as described by the Vaidya space-times has already been
analyzed in detail and strong curvature
naked singularities do form in such a collapse in generic situations,
either with or without the
geometric condition of self-similarity [2].

Hence, we need
to examine only the gravitational collapse of
type (i) matter fields for spherically symmetric space-times
considered here.
This is the form of  matter and
stress-energy for all  the observed fields so far with either
non-zero rest mass and also for zero rest mass fields, except the special
cases described by type (ii).
We take the matter fields to satisfy the weak energy condition,
i.e. the energy density as measured by any observer is non-negative
and for any timelike vector $V^a$
$$ T_{ab}V^aV^b\ge 0\eqn\qq$$
For the stress-energy tensor $T_b^a$ of type (i),
we can write
$$T^{ab}=\lambda_1 E^a_1E^b_1 +\lambda_2 E^a_2E^b_2
+\lambda_3 E^a_3E^b_3 +\lambda_4 E^a_4E^b_4\eqn\qq$$
where (${E_1,E_2,E_3,E_4}$) is an orthonormal basis with ${E_4}$
being timelike eigenvector and $\lambda_{i}$($ i=1,2,3,4$) are the
eigenvalues. For such a spherically symmetric matter distribution
we can choose coordinates $(x^i=t,r,\theta,\phi)$ to write the
metric as,
$$ds^2=-e^{2\nu}dt^2+e^{2\psi}dr^2+R^2d\Omega^2\eqn\qq$$
with $d\Omega^2= d\theta^2+ \sin^2\theta d\phi^2$ is the line element
on two-sphere.
Here $\nu,\psi$ and $R$ are functions of $t$ and $r$ and the stress energy
tensor $T^a_b$  given by equation (1)
has only diagonal components in this coordinate system (i.e.
we are using comoving coordinate system)
$$T^t_t=-\rho,\quad T^r_r=p_1,\quad T^{\theta}_{\theta}=p_2=
T^{\phi}_{\phi}=p_3,\quad T^t_r=T^r_t=0\eqn\qq$$
The quantities
$\rho, p_1$, $p_2$, and $p_3$ are the eigenvalues of $T^a_b$ and are
interpreted as the density and principle pressures.
Then, the weak energy condition holds for type (i)
matter fields provided,
$$\rho\ge0,\quad \rho+p_\alpha\ge0,\quad \alpha=1,2,3\eqn\qq$$

We note that  $R(t,r)\ge0$ here is the area coordinate
in the sense that the quantity $4\pi R^2(t,r)$ gives the proper area
of the mass shells and the area of such a shell at $r=const.$
goes to zero when $R(t,r)=0$. In this sense, the curve $R(t,r)=0$
describes the singularity in the space-time where the mass shells are
collapsing to a vanishing volume. Such a singularity is often called
a `shell-focusing'. It thus follows that the
range of the coordinates in metric (3) is given by
$$0\le r\le \infty,\quad -\infty<t< t_0(r)\eqn\qq$$
with $\theta$ and $\phi$ having the usual range of values.
The time $t= t_0(r)$ corresponds to the value of area coordinate
$R=R(t,r)=R(t_0(r),r)=0$
where the area of the shell of matter at a constant value of $r$
vanishes. This corresponds to the time when the matter shells meet the
physical singularity.

The Einstein field equations and the Bianchi identities
$T^a_{b;a}=0$ are written as below:
$$G^t_t=k_0T^t_t\Rightarrow
\left[R(-1+e^{-2\psi}R'^2-e^{-2\nu}\dot R^2)\right]'
=R^2R'k_0T^t_t\eqno(7a)$$
$$G^r_r=k_0T^r_r\Rightarrow
\dot{\left[R(-1+e^{-2\psi}R'^2-e^{-2\nu}\dot R^2)\right]}=R^2
\dot Rk_0T^r_r\eqno(7b)$$
$$G^t_r=k_0 T^t_r=0\Rightarrow \dot R'-\dot \psi R'-\nu '\dot R
=0\eqno(7c)$$
$$T^a_{t;a}=0\Rightarrow \dot T^t_t+
T^t_t(\dot \psi +{2\dot R\over R})-T^r_r\dot \psi=2p_2{\dot R
\over R}\eqno(7d )$$
$$T^a_{r;a}=0\Rightarrow (T^r_r)'+
T^r_r(\nu ' +{2R'\over R})-T^t_t\nu'=2p_2{R'
\over R}\eqno(7e )$$
where $k_0=8\pi G/c^2$ is the gravitational constant,
and in equations (7a) and (7b) equation (7c)
has been used. The $(')$ and
$\dot {( )}$ denote partial derivatives with respect to $r$ and $t$.
As we consider the collapse problem, we take
$\dot R<0$.
Eliminating $p_2$ from equations (7d), (7e) and using (7c) imply
$$ {\partial\over\partial r}[T^r_rR^2\dot R] - {\partial\over\partial t}
[T^t_tR^2R']=0$$
and we conclude That
$$ T^t_t=-\rho=-{F'\over k_0R^2R'},\quad
T^r_r=p_1=-{\dot F\over k_0R^2 \dot R}\eqno(8)$$
where $F=F(t,r)$ is an arbitrary function of $t$ and $r$.
For the Tolman-Bondi dust collapse [8] or perfect fluid space-times [9],
$F$ is physically interpreted as mass function.
For dust collapse, $F=F(r)$
represents the total mass within a coordinate radius $r$.
Thus, the function
$F$ is treated as the mass function for the cloud with $F\ge0$.
In general, $R'$ may not be positive, however, in cases when
$R'\ge0$, the weak energy condition implies from (8) that
$F'\ge0$. In the case of gravitational collapse
$(\dot R<0)$, it will be seen
later from the null geodesic equations, that $R'<0$ at the singularity
implies that no rays will be outgoing and the singularity will
be censored.

Using (8), equations (7d) and (7c) become
$$4p_2R\dot RR'=
-2\dot F'+F'{\dot G\over G}+\dot F{H'\over H}\eqno(9a)$$
$$-2\dot R'+R'{\dot G\over G}+\dot R{H'\over H}=0\eqno(9b)$$
where we have put
$$G=G(t,r)=e^{-2\psi}(R')^2,\quad
H=H(t,r)=e^{-2\nu}\dot R^2,
\eqno(9c)$$

Integration of the remaining
field equations $G^t_t=k_0T^t_t,\quad G^r_r=k_0T^r_r$ is
straight forward and some simplification gives
$$ \rho={1\over k_0R^2}(F,_R+{F,_r\over R'}),\quad p_1=-{F,_R\over
k_0R^2}\eqno(10a)$$
$$(F,_{RR}+2k_0p_2R)T\equiv {T\over p}=
F,_r{G,_R\over 2G}-F,_{rR}\eqno(10b)$$
$$-2T,_R+T{G,_R\over G}+T{H,_R\over H}+{H,_r\over H}=0$$
\noindent
$$\Rightarrow
-2[p(F,_r{G,_R\over 2G}-F,_{rR})],_R+
p(F,_r{G,_R\over 2G}-F,_{rR})({G,_R\over G}+{(G-1+{F\over R}),_R\over
(G-1+{F\over R})})$$
\noindent
$$={G,_r+{F,_r\over R}\over
G-1+{F\over R}}\eqno(10c )$$
$$ H=G-1+{F\over R}\eqno(10d)$$
We have used $R$ instead of $t$ as variable in the above
equations. The function $p=p(R,r)$ is defined in  (10b),
$F(t,r)=F(R,r)$, $T(R,r)=R'$ and
likewise $G(R,r)$ and $H(R,r)$ are to be treated
as functions
of $R$ and $r$. Here ($,_R $) and ($,_r $) denote
partial derivatives with respect to
$R$ and $r$ and are defined by
$$\left[{\partial \over \partial r}\right]_{t=const.}
=R'\left[{\partial \over \partial R}\right]_{r=const.}
+\left[{\partial \over \partial r}\right]_{R=const.}$$
$$\left[{\partial \over \partial t}\right]_{r=const.}=
\dot R \left[{\partial \over \partial R}\right]_{r=const.}
\eqno(11)$$
The two equations in (10c) are equivalent. The later
equation of (10c) is obtained from the former by the substitution of
$T$ from (10b).
The remaining field equation $G^{\theta}_{\theta}=k_0T^{\theta}_{\theta}=
G^{\phi}_{\phi}=k_0T^{\phi}_{\phi}$ is then just a consequence of
(10a) to (10d).

In all we have five unknowns, namely $T(R,r),G(R,r),H(R,r)$, $F(R,r)$ and
$p_2(R,r)$, and three equations (10b), (10c), and (10d) relating them.
In fact, the functions
$F$ and $p_2$ determine the form of matter and
the equation of state one is dealing
with. For example, for dust models $p_2=0=F,_R$ which implies
$F(R,r)\equiv F(r)$, and for a perfect fluid
$p_2=-F,_R/R^2$, if the fluid has the equation of state $\rho+p=0$.
In addition then $F,_r=0$ implies
$F(R,r)\equiv F(R)$.
Therefore, one starts with a particular
stress-energy tensor by selecting these two
functions and then the geometry of
space-time or the metric functions are
determined as follows. Knowing $F(R,r)$ and $p_2(R,r)$
one determines
$G=G(R,r)$ from (10c), which is a second order partial differential
equation, with appropriate initial and boundary data for $G$.
Note that the equation (10c) is a parabolic type (in other words
a generalized heat wave type equation) second order partial
differential equation and as such could be solved with a possible set of
initial and boundary conditions given by $G(R,0)=g(R),G(a,r)=g_1(r)$
and $G(b,r)=g_2(r)$.
Equation (10b) then determines $R'=T(R,r)$ as a function of $R$ and $r$,
which on integration yields
$R=R(t,r)$. The function $H=H(R,r)$ is immediate from (10d).
In case $F,_{RR}+2k_0p_2R\equiv 1/ p=0$
identically, then (10b) breaks down and one cannot
determine $R'=T(R,r)$ from (10b). Rather it implies $(G/F,_r),_R=0$, thus
determining $G(R,r)$ instead of $T(R,r)$. The function $T(R,r)$ in
such cases is  determined by
integrating (10c) with appropriate
initial conditions by treating it as an ordinary differential
equation for $T$ in variable $R$ and specifying initial conditions
accordingly.

\beginsection {3. THE INITIAL VALUE PROBLEM}

Consider now  a spherically symmetric cloud of
matter collapsing
gravitationally to give rise to a space-time singularity
$R=0$ at the center $r=0$. Our problem is to characterize the
conditions under which this  could be naked, and those in
which the singularity is completely covered by an event
horizon formed during
the collapse. Thus, for example, for
a homogeneous gravitational
collapse of dust described by the Oppenheimer-Snyder models,
the resulting singularity is fully covered by an
event horizon. On the other hand, if inhomogeneities are
present, strong curvature naked singularities do form in such
collapse scenarios (see e.g. [2]).

We define  regular initial
data on a spacelike hypersurface $t=t_i$
from which the collapse starts. On the
surface  $t=t_i$ we require physical quantities
such as density, pressures etc. be non-singular.
Matter has a compact
support on $t=t_i$ and $r=r_b$ denotes the boundary of the
object. One requires appropriate boundary conditions to match
the interior metric of the cloud to the exterior.
The exact boundary conditions will depend
on what the exterior spacetime is,
which could be vacuum Schwarzschild or a radiating Vaidya metric etc.

The space-time singularity occurs at
the time $t=t_0(r)$, which corresponds to  $R(t,r)=0$.
Let $t=t_0(0)>t_i$
be the first point of the singularity, which is the time
of the singularity occurring at $r=0$. (If $t=t_0(0)$
is not already the first point of the singularity curve $t=t_0(r)$,
it could be made so by a simple translation of the coordinate $r$.)
Thus, this implies
a boundary condition that for $t_i< t<t_0(0)$, the center
of the cloud $r=0$ is a regular center. In terms of the functions above
defining the gravitational collapse, this amounts to the
requirement,
$${F,r\over R'}< \infty\eqno(12)$$
at the regular center $r=0$ for $t_i< t<t_0(0)$.
We are interested in analyzing the nature of this first shell-focusing
singularity $R=0$ which we call a central singularity when it occurs
at $r=0$. Thus, we assume that
there is a neighborhood of the central singularity such that $R'>0$
for $r>0$.

Basically we would  require some general differentiability
conditions for functions $F$, $p_2$ and $R$ and take $F(R,r)$ and
$p_2(R,r)$ to be at least $C^2$ for $R>0$
and  $R(t,r)$ as $C^2$ for all $t$ and $r$.

As mentioned earlier, the choice of a particular
matter distribution is made by selecting the physical quantities
$F$ and $p_2$. Since the data at the initial hypersurface is non-singular,
this puts restrictions on the values of these
functions at the initial surface  $t=t_i$. That is, one has
to choose $F$ such that at the initial surface
$F,_r/R^2R'<\infty$, $F,_R/R^2<\infty$ and $p_2<\infty$.
Therefore, the condition that initial data be non-singular
means a proper choice of free functions $F$ and $p_2$.

To make this clear,
one could use the
coordinate freedom left in the rescaling of the coordinates
$r$ and $t$
without any loss of
generality, so that at $t=t_i$,
$$R(t_i,r)=r\eqno(13)$$
At $t=t_i$ the quantities $\rho, p_1,
p_2 $ etc. do not
diverge at $r\ge0$.
Since the initial data is to be non-singular at $r=0$
on this surface, the
first derivatives of $F$ must have the
following behavior at $r=0$,
$$[(F_{,r})_{R=r}/r^2]_{r=0}<\infty,
\quad [(F_{,R})_{R=r}/r^2]_{r=0}<\infty,
\quad [p_2(r,r)]_{r=0}< \infty\eqno(14)$$

Similarly, requiring that $G(R,r)<\infty$ at this $R=r$ surface
at $r=0$ puts restriction on the choice of $p_2$ and
second derivatives
of $F$. That is, from equation (10b) one has
$$\left[{[F_{,RR}]_{R=r}+2k_0p_2(r,r)r+ [F_{,rR}]_{R=r}
\over [F_{,r}]_{R=r}}
\right]_{r=0}\ne - \infty\eqno(15)$$
at $t=t_i$ and therefore
$e^{2\psi}=1/G\ne \infty$. This ensures that the initial data
is non-singular at $t=t_i$. We do not discuss  further
implications of the conditions such as above but the
point is the choice of a non-singular initial data  and
boundary conditions restrict the functions
$F$ and $p_2$ suitably ensuring their proper
choice.

The space-time singularity appears at the point $R=0,r=0$, and
therefore the behavior of various functions near the singularity is
important. To examine this, we
get after some simplification from equations (10a) to (10d),
$$ \rho={1\over k_0u^2X^2}(\Lambda,_X+{\eta\over \beta}),\quad p_1=-
{\Lambda,_X
\over
k_0u^2X^2},\quad p_2=p_2(X,u)\eqno(16a)$$
$$ \beta(X,u)
 =P(\eta{f,_X\over 2(1+f)}-\eta,_X)\eqno(16b)$$
$$-2[{P\over \sqrt{1+f}}(\eta{f,_X\over 2(1+f)}-\eta,_X)],_X\sqrt{1+f}+
{(f+{\Lambda\over X}),_X\over
f+{\Lambda\over X}}[P(\eta{f,_X\over 2(1+f)}-\eta,_X)-X]$$
\noindent
$$=
u{f,_u+{\Lambda,_u \over X}\over f+{\Lambda\over X}}\eqno(16c)$$
$$ H=f+{\Lambda\over X}\eqno(16d)$$
where we put for the sake of convenience $G(X,u)\equiv 1+f(X,u)$
and  introduce two variables
$u=r^{\alpha}$ ($\alpha \ge 1$ is a constant), $X=R/r^{\alpha}=R/u$
and the following notation:
$$\eta =\eta(X,u)={F,_r\over r^{\alpha-1}},\quad \Lambda =\Lambda (X,u)
={F(R,r)\over r^{
\alpha}}={F\over u}$$
$$P=P(X,u)={1\over \Lambda,_{XX}+
2k_0p_2u^2X},\quad \beta(X,u)={R'\over r^{\alpha-1}}
={T\over r^{\alpha-1}}
\eqno(17)$$
Here $G,H,\eta,\Lambda$ and $\beta $
are all functions of $X=R/u$ and $u=r^{\alpha}$.
The constant  $\alpha \ge 1$ is to be chosen so that
$\beta=T/r^{\alpha-1}$
does not vanish or go
to infinity identically as  $r\to 0$ in the  limit of approach to the
singularity
along all $X=const.$ directions. Here
$(,_X)\quad (,_u)$ represent partial
derivatives with respect to $X$ and $u$
respectively.
The weak energy condition implies for these functions,
$$ {\eta\over \beta }\ge 0,\quad \Lambda, _X+{\eta\over \beta}\ge 0,\quad
k_0u^2X^2p_2+\Lambda, _X+{\eta\over \beta}\ge 0
\eqno(18)$$
Equation (16c)
is a second order partial differential equation for $G(X,u)=1+f$
and is solved by specifying the initial and boundary conditions on
$G(X,u)$. Knowing $G$ rest of the unknowns are then immediate from
equations (16a) and (16d).

The values $G(X_0,0)=1+f(X_0,0),\quad \beta(X_0,0)$  for some positive
$X=X_0$ are of significance
in our analysis. The function
$\beta(X,0)$ can be calculated from (16b) once
$f(X,0)$ is fixed except as
mentioned earlier in the case where $\Lambda,_{XX}+
2k_0p_2u^2X=0$ identically. In such a
case the function $\beta(X,0)$ could not be calculated from (16b),
rather it implies
$1+f(X,u)=\eta g(u)$ where $g(u)$ is an arbitrary function of $u$.
Thus $f(X,u)$ is determined from  (16b) instead of $\beta$. Then
$f(X,0)$
is again chosen (by selecting $g(u)$) and $\beta(X,u)$ is then
determined by integrating an ordinary
differential equation (knowing $G=1+f$ actually  (10c) becomes
an equation for $\beta$) to find $\beta(X,u)$ given by
$$-2\beta,_X+\beta {f,_X\over 1+f}+(\beta-X){(f+{\Lambda\over X}),_X
\over
f+{\Lambda\over X}}$$
\noindent
$$=u{f,_u+{\Lambda,_u\over X}\over
f+{\Lambda\over X}}\eqno(19)$$
Now, $\beta(X,0)$ can be determined from  integration of the above.

The initial values of $G(X,u)\equiv 1+f$ and $\beta(X,u)$
at $X=X_0,u=0$ have  to be chosen
while selecting a particular model of
spherically symmetric gravitational
collapse and this has to be a physically reasonable
choice. In fact (16c), which determines $f$ is a parabolic
type second order partial differential equation and as such
could be solved with
a given set of initial data which could be of the form
$f(X,0), f(a,u)$ and $f(b,u)$. Similarly in the special case when
$1/P=0$ identically, one can choose $\beta(X_0,0)$  as part of the
initial condition to solve the ordinary differential equation determining
$\beta(X,u)$. Hence the spacetime subject to the chosen form of matter
is determined by the solutions of either the second order
parabolic partial differential equation (16c) or the second order
ordinary differential equation (19) as the case may be.
The input to these equations come from the prechosen form of the matter
by the way of functions $P, \Lambda$and $\eta$( i.e. $F$ and $p_2$).
The initial data is given as a set of appropriate values $a,b,c,\lambda,
\mu$ for $X,u,f, f_X,f_u$ in equation (16c) and $a_1,b_1,\beta_1,
\beta_2$ for $X,u, \beta, \beta_X$ in case of equation (19). The
existence of solutions of such equations (16c) or (19) (as the case may be)
has been studied and these have been shown to exist with a suitable
arbitrary choice of initial and boundary data under fairly general
condition. In fact, under fairly general conditions solutions exist
with arbitrary choice of $f(X,0)$ and $f_X(X,0)$ in case of the
parabolic partial differential equation (16c) and similarly with
arbitrary choice of $\beta(X_0,0)$ in case of ordinary differential
equation (19).

Therefore, the point is  that  for a given form of matter (i.e.
given $F$ and $p_2$), the
values of $\beta(X_0,0)$ (or $f_{,X}(X_0,0)$ and $f(X_0,0)$
at some positive value of
$X=X_0$ are part of the initial data and the solution to field equations
with these initial values represents one of the spacetimes with the
above specified form of matter.
Hence, one can make a suitable choice of
these functions.
One such reasonable choice could be as follows. If
the usual Lorentz-Minkowskian geometry is to be valid in an infinitesimal
neighborhood of the regular center $r=0$, then one must require that the
circumference $2\pi R$ of an infinitesimal sphere about the center be
just the $2\pi$ times its proper radius $e^{\psi}dr$. In other words,
$$e^{2\psi}=(R')^2\quad {\hbox{at}}\quad r=0\Rightarrow G=1 \quad
{\hbox{at}}\quad
r=0 \eqno(20)$$
Hence, $f(X)=G(X,0)-1=0$
is one of the many possible initial conditions.

\beginsection {4. THE EXISTENCE OF NAKED SINGULARITY}

The existence of a naked singularity in  space-time is
characterized by the presence of outgoing families of future directed
non-spacelike geodesics, which are past incomplete
and terminate in the past at the singularity.

Radial null geodesics for a spherically symmetric space-time (3)
are given by $ds^2=0$,
$${dt\over dr}={{dt/ dk}\over {dr/ dk}}=
{K^t\over K^r}=e^{\psi-\nu}\eqno(21)$$
$${d\over dk} [e^{\psi}K^r] +e^{2\psi}(K^r)^2[e^{-\nu}
\dot \psi-e^{-\psi}\nu']
=0\eqno(22)$$
where $K^t$ and $K^r$ are the only non-zero
components of the tangent vector $K^a$($K^{\theta}=0,K^{\phi}=0$) and
$k$ is an affine parameter along the null geodesics.
The singularity appears at the point
$R(t,r)=0,r=0$, therefore if there are outgoing future
directed radial null geodesics terminating at the singularity in past,
then $R\to 0$ as $r\to 0$ along these geodesics.

We have
from the above,
$${dR\over du}= {1\over \alpha r^{\alpha -1}}(
R,_r+R,_t{dt\over dr})={(1-{\Lambda\over X})\sqrt {G}\over
(\sqrt{G}+\sqrt{H})T}=(1-\sqrt{{f+{\Lambda\over X}\over 1+f}})\beta\equiv
U(X,u)\eqno(23)$$
Note that in the case of collapse, since $\dot R<0$, $dR/du$
becomes negative if $R'<0$.
Hence, in such a case
the geodesics are all ingoing and the singularity is
censored.
For an outgoing geodesics $dR/du$ must be positive and hence we require
that $R'\ge0$ at the singularity. It further follows
that if $dR/du$ is negative ( for $R<F$,  $dR/du$ is negative)
geodesics become ingoing (in the sense that area coordinate $R$ starts
decreasing).
Note that $F(R,0) > 0 $ implies
$dR/du \to -\infty$  at the singularity and so the geodesics are
all ingoing, which corresponds to a Schwarzschild type situation
where mass is already present at the center $r=0$. When $F=0$ at the
first point of the singularity, the situation may correspond either to a
black hole or a naked singularity.
For example, in homogeneous dust collapse,
$F\propto r^3$ and $F=0$ at the first point which is covered by horizon.
The point $R=0,u=0$ is
a singularity of the differential equation (24) and hence in order to
determine whether geodesics do terminate at the singularity or not
one has to analyze the behavior of  characteristic curves in the
vicinity of the singular point. If  radial null geodesics do terminate
at the singularity then we have
$$ X_0=\lim_{R\to 0,u\to 0}\left({R\over u}\right)=
\lim_{R\to 0,u\to 0}\left({dR\over du}\right)=U(X_0,0)\eqno(24 )$$
If a real and positive value of $X_0$ satisfies the above equation then
the singularity could be naked. On the other hand, if the
above has no real positive
roots, clearly the singularity is not
naked with no families of non-spacelike trajectories coming out.
Therefore, the necessary condition for the singularity to be naked
is   $V(X)=0$ has a real positive root $X=X_0$ where
$$V(X)\equiv \left(1-\sqrt{{f(X,0)+{\Lambda(X,0)\over X}\over 1+f(X,0)
}}\right)\beta(X,0)
-X=0\eqno(25)$$

As pointed out earlier, one could select
$G(X,0)=1+f(X,0)$
as an initial data and  rest of the unknowns in the above
equation, namely $\beta(X,0)>0$ is implied by the field equations.
That $V(X)=0$ has a real positive root is a necessary
condition for the singularity to be naked, but need not be
a sufficient condition. To examine this, consider the equation of
radial null geodesics in the form
$u=u(X)$ given by
$${dX\over du}={1\over u}\left({dR\over du}-X\right)={U(X,u)-X\over u}
\eqno(26)$$
Integration of the above  yields  radial null
geodesics in the form $u=u(X)$. Let $X=X_0$ be a simple real positive
root of $V(X)=0$. If geodesics are to terminate at the singularity
$R=0,u=0$ then $u\to 0$ as $X\to X_0$ along the same. We could
then decompose $V(X)$ as
$$V(X)\equiv (X-X_0)(h_0-1)+h(X)\eqno(27)$$
where $h(X)$ is chosen such that it contains higher order terms in $X-X_0$
$$h(X_0)=\left[{dh\over dX}\right]_{X=X_0}=0 \eqno(28)$$
Using equations (27), (16b) to (16d), (17), and (28) we get
for $h_0$
$$h_0={(\beta_0-1)(\beta_0 -1-X_0)\over \beta_0 -X_0}(\beta_0P_0+X_0)
-{1+X_0\over 2X_0^2V_0(\beta_0-X_0)}(\Lambda_0+X_0
q_0)\eqno(29)$$
where
$$\beta_0=\beta(X_0,0),\quad \Lambda_0=
\Lambda(X_0,0),P_0=P(X_0,0)\eqno(30)$$
$$q_0=\left[\Lambda,_X\right]_{u=0,X=X_0}\eqno(31)$$
Writing $S=S(X,u)=U(X,u)-U(X,0)+h(X)$, we could write (26) as
$${dX\over du}-(X-X_0){h_0-1\over u}={S\over u}\eqno(32)$$
Note that because of the way $S(X,u)$ is defined $S(X_0,0)=0$, i.e.
in the limit $u\to 0,X\to X_0$ we have $S\to 0$. Integration of
the above is straight forward by multiplication of an integrating factor
$u^{-h_0+1}$ and we get
$$X-X_0=Du^{h_0-1}+u^{h_0-1}\int{Su^{-h_0+1}du}\eqno(33)$$
Here $D$ is a constant which labels different geodesics. If geodesics
described by the above equation do terminate at the singularity,
$u\to 0$ as $X=X_0$ in the above. To see this, note that as
$X\to X_0, u\to 0$ the last term of the above  always vanishes
near the singularity since $S\to 0$ as $u\to 0,X\to X_0$. The first
term, i.e. $Du^{\alpha-1}$ vanishes only
if $h_0>1$. It follows  that the integral curve
(radial null geodesic)
$D=0$ always terminates at the
singularity $R=0,u=0$ with $X=X_0$ as tangent. Further,
if $h_0>1$, a family
of outgoing radial null geodesics  terminates at the singularity
in past, each curve given by a different value of the constant $D$.

It follows that if $V(X)=0$ has a real positive root then the
gravitational collapse would terminate in a singularity which would
at least be locally naked.
As we have discussed earlier, such a condition basically corresponds
to the choice of initial data for the differential equation (16c) in
the form of the choice of $f(X_0)=G(X_0,0)-1$, or in the choice of
$\beta(X_0,0)$ in (19) as the case may be.
Our analysis here implies that for all the presently known naked
singular spherically symmetric  examples with equations of state such as
dust or perfect fluid etc., there is a similar naked singular spacetime
for all nearby equations of state in the sense defined by the existence
of solutions of the parabolic differential equation discussed above.
It is thus clear that for a wide range of spherically symmetric
gravitational collapse,
irrespective of the form of the matter,
or a particular equation of state, a naked
singularity would form in the above sense.

Next, we determine  the curvature strength of the naked singularity.
This is determined in terms of the curvature growth in the limit of
approach to the naked singularity.

Consider the scalar quantity
$$\Psi= R_{ab}K^aK^b\eqno(34 )$$
For the space-times (3), using (21) and
(22) and the fact
that $K^a$ is a null vector,
we get
$$\Psi=T^a_bK^bK_a=T^t_tK^tK_t+T^r_rK^rK_r=\left[{F'\over R'}-
{\dot F\over \dot R}\right]{e^{2\psi}(K^r)^2\over R^2}=
{\eta  e^{2\psi}(K^r)^2\over TR^2}\eqno(35)$$

The singularity is said to be a strong curvature singularity [10] if
$$\lim_{k\rightarrow 0}k^2\Psi\neq 0\eqno(36)$$
We therefore have
$$\lim_{k\rightarrow 0}k^2\Psi=\lim_{k\rightarrow 0}
{k^2\eta e^{2\psi}(K^r)^2\over \beta R^2}\eqno(37)$$
Using the equations above and l'Hospitals rule  we get
$$\lim_{k\rightarrow 0}k^2\Psi\propto {\eta_0\over \beta_0^2}\eqno(38)$$
Therefore, as long as $\eta_0 \neq 0$ the strong curvature condition
is satisfied. In other words,
in all the situations  the singularity would be
strong if the energy
density (i.e. $\rho + p_1=\eta/\beta u^2X^2$) does not
vanish in the neighborhood of the singularity.

\beginsection {5. CONCLUDING REMARKS}

In this paper we have  shown that the phenomena of naked singularity
is dependent on the initial values chosen for
solving the field equations, in that for all sets of
regular initial values which produce
at least one positive root of the equation $V(X)=0$,
the singularity would be naked.

It follows that a naked singularity could
develop in a generic situation involving spherically symmetric collapse
of matter from non-singular initial data. Therefore,
in order to preserve the cosmic censorship hypothesis one has
to avoid all such initial data and hence a deeper analysis of
equation (25) is required in order to determine such initial data
and the kind of physical parameters they would specify.
This would, in other words, classify the
range of physical parameters to be avoided for a particular form of
matter. More importantly, it would also pave the way for the black hole
physics to use only those ranges of allowed parameter values which
would produce
black holes, thus putting the black hole physics on a more reasonable
footing.

\vfil\eject

\centerline{REFERENCES}

\item{\bf 1.} P. S. Joshi and I. H. Dwivedi, Commun. Math. Phys.
146, p.333 (1992); Lett. Math. Phys. 27, 235 (1993).
\item{\bf 2.} K. Lake, Phys. Rev. Lett. 68, 3129 (1992);
P. S. Joshi and I. H. Dwivedi, Phys. Rev. D47, 5357 (1993).
\item{\bf 3.} I. H. Dwivedi and P. S. Joshi, Class. Quantum Grav.
6, 1599 (1989); 8, 1339 (1991).
\item{\bf 4.} B. Waugh and K. Lake, Phys. Rev. D38, 4, 1315 (1988);
A. Ori and T. Piran, Phys. Rev. D42, 1068 (1990).
\item{\bf 5.} P. S. Joshi and I. H. Dwivedi, Gen. Relat.
Grav. 24, 129 (1992); Phys. Rev. D45, 2147 (1992);
J. P. S. Lemos, Phys. Rev. Lett. 68, 1447 (1992); P. Szekeres and
V. Iyer, Phys. Rev. D47, 4362 (1993).
\item{\bf 6.} D. M. Eardley in "Gravitation in Astrophysics", Plenum
Publishing Corporation Edited by B. Carter and J. B. Hartle 223 (1987).
\item{\bf 7.} D. Kramer et al. `Exact solutions of Einstein's
field equations, Cambridge University Press, Cambridge (1980).
\item{\bf 8.} H. Bondi, Mon. Not. Astron. Soc. 107, p.343 (1947);
\item{\bf 9.} C. W. Misner and D. H. Sharp, 136, B571 (1964).
\item{\bf 10.} F.J.Tipler, C.J.S.Clarke and G.F.R.Ellis,
"General Relativity and Gravitation", Vol 2, (ed.A.Held),
New York: Plenum Press, 97 (1980).

\end